\documentclass[letterpaper, 10 pt, conference]{ieeeconf}  

\IEEEoverridecommandlockouts                              

\makeatletter  
\let\NAT@parse\undefined
\makeatother
\usepackage[hidelinks]{hyperref}

\usepackage[caption=false]{subfig}
\usepackage{amsfonts} 
\usepackage{mathtools} 
\usepackage{graphicx}
\usepackage{tikz}
\usepackage[ruled,vlined,linesnumbered]{algorithm2e}  
\usepackage{etoolbox}

\usepackage{amsthm}

\newcommand{\T}{^{\mathsf{T}}}

\newcommand{\B}[1]{\if#1\relax\bm
{#1}\else\mathbf{#1}\fi} 
\newcommand{\R}[1]{\mathrm{#1}}						      
\newcommand{\C}[1]{\mathcal{#1}}
\newcommand{\BB}[1]{\mathbb{#1}}
\newcommand{\norm}[1]{\left\lVert #1 \right\rVert}
\newcommand{\abs}[1]{\left\lvert #1 \right\rvert}
\newcommand{\restr}[2]{\left. #1 \right\rvert_{#2}}

\makeatletter
\patchcmd{\@algocf@start}
  {-1.5em}
  {0pt}
  {}{}
\makeatother

\SetKwComment{tcc}{ {$\triangleright$} }{}  
\SetKwComment{tcp}{ {$\triangleright$} }{}  

\SetCommentSty{myCommentStyle}

\graphicspath{{figures/}}  
\allowdisplaybreaks        

\title{Data-driven architecture to encode information in the kinematics of robots and artificial avatars}

\author{Francesco De Lellis\textsuperscript{1}, Marco Coraggio\textsuperscript{2}, Nathan C. Foster\textsuperscript{3}, Riccardo Villa\textsuperscript{3},\\ Cristina Becchio\textsuperscript{3}, Mario di Bernardo\textsuperscript{1,2}
\thanks{\textsuperscript{1}Department of Electrical Engineering and
Information Technology, University of Naples Federico II, Naples, Italy}%
\thanks{\textsuperscript{2}Scuola Superiore Meridionale, Naples, Italy}%
\thanks{\textsuperscript{3}Department of Neurology, University Medical Center Hamburg-Eppendorf, Hamburg, Germany}%
\thanks{This work was supported by the EU Research Project SHARESPACE (EU HORIZON-CL4-2022-HUMAN-01-14). For more info see {\tt www.sharespace.eu}}
}

\newtheorem{lemma}{Lemma}
\newtheorem{assumption}{Assumption}

\begin{document}

\maketitle

\begin{abstract}
We present a data-driven control architecture for modifying the kinematics of robots and artificial avatars to encode specific information such as the presence or not of an emotion in the movements of an avatar or robot driven by a human operator. We validate our approach on an experimental dataset obtained during the reach-to-grasp phase of a pick-and-place task.
\end{abstract}

\section{Introduction}

Movement encodes significant information about both the external characteristics of objects and the internal states of the mover, such as intentions and expectations \cite{becchio2018seeing}. For instance, the way an individual reaches towards an object can reveal 
mover's anticipations regarding its weight \cite{podda2017heaviness}, and their specific intentions, such as whether they plan to pour or drink from it \cite{scaliti_kinematic_2023}. Thus, analyzing movement patterns allows the inference of others' internal states \cite{becchio2018seeing, gallivan2018decision, wispinski2020models}. However, decoding or reading this information accurately is challenging due to the variability in movement kinematics and the observer's ability to distinguish between informative and non-informative variations.

Experimental studies have shown that naive human observers can decode or readout some but not all information from movement kinematics \cite{becchio2018seeing}, with the potential to overlook informative variations \cite{scaliti_kinematic_2023} or misinterpret the data \cite{montobbio2022intersecting}. During social interactions, individuals naturally modify their movement kinematics to make their actions more interpretable by others \cite{mcellin2018identifying, strachan2021evaluating}, highlighting the importance of conveying information clearly through movement. Consequently, for robots and virtual reality (VR) avatars to improve interactions with humans, it is essential to accurately encode information in their kinematics, making the encoded information more accessible and interpretable.

In this Letter, we present a data-driven control architecture aimed at manipulating information within robot and artificial avatar kinematics—either by encoding it. After reviewing current advancements, we delve into the mathematical basis of our approach, particularly within the context of the pick-and-place task, a paradigmatic problem in social robotics \cite{mortl2014rhythm}. Our focus is on adjusting the reach-to-grasp movement kinematics of an avatar driven by a human in VR to encode emotional states, such as fear. The proposed architecture enables real-time dynamic adjustments to the motion representation, combining live kinematics with a database of movements using AI tools, ensuring that avatars or tele-operated robots in Extended Reality (XR) facilitate the encoding of information.

\subsection{State of the art}
Emotion, as a relatively abstract concept, has been integrated into dynamical and control systems in diverse ways, reflecting its complexity. This integration spans decision-making architectures and attempts at mimicking cognitive processes, illustrating the variety of approaches in current research. In decision-making, a framework was developed in  \cite{huang2021connecting} combining model-based control and model-free reinforcement learning, informed by cognitive science insights into emotional responses. Similarly, emotion's role in mimicking cognitive processes is evident in reinforcement learning strategies for recurrent neural network parameter tuning presented in  \cite{huang2018braininspired}, where emotional state-modulated reward functions enhance learning and control.

Efforts to interact with humans have led to socially aware robotic systems that communicate emotions, using kinematic redundancy to encode emotions in movements \cite{claret2017exploiting}, and trajectory planning that incorporates emotional aspects, drawing on Laban Movement Analysis \cite{lourens2010communicating, bernardet2019assessing, wu2022robotic}. Moreover, VR has been used to study emotion encoding in body kinematics, minimizing emotion misclassification in exergame scenarios \cite{lombardi2021dynamic}. Studies have shown that human observers can recognize emotions from body movements, suggesting the feasibility of communicating emotions through motion \cite{atkinson_emotion_2004, llobera2022playing}.

Despite progress, defining emotion concisely remains challenging, with research ongoing into quantifying how human movement conveys emotions and social intentions \cite{melzer_how_2019, 10.3389/frobt.2020.532279}. Addressing these challenges, this paper proposes a trainable architecture for encoding desired emotions in movement kinematics from human demonstrations and a tunable reinforcement learning strategy with stability guarantees for controlling end-effector positions, advancing emotion-aware control systems.

\section{Control of body movement to express emotion}
\label{sec:control_design}

Given a movement by a person, say $H$, another human or computer-based evaluator, say $L$, can observe such movement and decide whether $H$ was experiencing or not a certain emotion while performing it.
If $H$ was actually experiencing the emotion \emph{and} this was successfully recognized by $L$, we say that the movement \emph{encodes} the emotion, and that $L$ could \emph{decode} the movement.
Given a human movement not encoding a certain emotion, we consider the problem of altering it to produce a motion that encodes the desired emotion (or vice-versa that does not if the original motion does). Such a motion could then be used as a reference motion for an avatar or robot tele-operated by the human so as for it to express (or not) the target emotion. 

Before formalizing the problem mathematically, we review a few useful definitions and notation.

\subsection{Preliminaries}

\paragraph*{Notation}
When applied to vectors, $\norm{\cdot}$ denotes the Euclidean norm.
$\odot$ denotes the Hadamard product.
$\C{B}^\C{A}$ denotes the set of functions from set $\C{A}$ to set $\C{B}$.

Let $T \in \BB{N}_{>0}$ be the \emph{(maximum) duration} of the movements being considered, 
$\C{T} \coloneqq \{1, \dots, T\}$ be the corresponding \emph{time window}, 
$\Delta t \in \BB{R}_{>0}$ be a \emph{sampling time},
and $\rho \in \BB{N}_{>0}$ be the \emph{number of degrees of freedom} being considered (e.g., $\rho=3$ if a single point is being sensed in the 3D space).
Let $\C{X} \coloneqq (\BB{R}^{\rho})^\C{T}$ be the sets of movement signals considered in particular, letting $p \in \C{X}$ and $v \in \C{X}$ be the position and velocity associated to the same movement, the constraint holds that 
\begin{equation}\label{eq:relation_position_velocity}
    v(t) = \frac{p(t+1) - p(t)}{\Delta t}, \quad \forall t \in \{1, \dots, T-1\}.
\end{equation}
Moreover, we assume that all movements eventually stop; hence, we let $\delta_{\R{vel}} \in \BB{R}_{> 0}$ be a small threshold, and define the \emph{terminal instant} associated to a movement with velocity $v$ as the time instant $t_{\R{term}}(v) \in \C{T}$ such that $\norm{v(t)} \le \delta_{\R{vel}}$ for all $t \ge t_{\R{term}}(v)$ 

We let $\R{dist} : \C{X} \times \C{X} \rightarrow \BB{R}_{\ge 0}$ denote the \emph{distance} between two signals, induced by the $L^2$ norm, that is
\begin{equation*}
    \R{dist}(x_1, x_2) \coloneqq \sqrt{\sum_{t = 1}^T\norm{x_1(t) - x_2(t)}^2}.
\end{equation*}

Moreover, given a subset of signals $\C{S} \subseteq \C{X}$, we define the \emph{$\C{S}$-projection} operator $\R{proj}_{\C{S}} : \C{X} \rightarrow \C{S}$, as%
\footnote{The time needed to compute $\R{proj}_{\C{S}}$ in \eqref{eq:projector} grows at most linearly with respect to the cardinality of $\C{S}$.}
\begin{equation}\label{eq:projector}
    \R{proj}_\C{S}(x_\C{X}) \coloneqq \arg \min_{x_\C{S} \in \C{S}} \R{dist}(x_\C{X}, x_\C{S}).
\end{equation}

\subsection{Problem statement}

We denote by $p_\R{h}, p_\R{a} \in \C{X}$ and $v_\R{h}, v_\R{a} \in \C{X}$ the position and velocity signals associated to the \emph{original human motion} to be modified and its \emph{altered} version, respectively.

The \emph{encoding function} $\varepsilon : \C{X} \rightarrow \{0, 1\}$ associates to a velocity signal the \emph{encoding level} of that movement, that is $1$ if the emotion is encoded and $0$ otherwise (note that $\varepsilon$ corresponds to an \emph{encoding model}, as described in \cite{turri_decoding_2022}).
Finally, we let $e_\R{des} \in \{0, 1\}$ be a desired value for the encoding level.
We aim to solve the optimization problem
\begin{subequations}\label{eq:problem_statement}
    \begin{align}
        \min_{v_\R{a} \in \C{X}} \ \ & \R{dist}(v_\R{h}, v_\R{a}),\\ 
        \text{s.t.} \ \
        & \varepsilon(v_\R{a}) = e_\R{des}.\label{eq:constraint_encoding}
    \end{align}
\end{subequations}

\begin{figure}[t]
    \centering
    \includegraphics[width=\columnwidth]{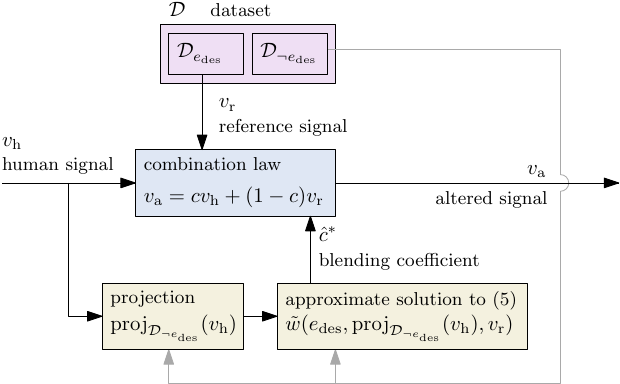}
    \caption{Block scheme of the proposed strategy to solve Problem \ref{eq:problem_statement}.}
    \label{fig:block_scheme}
\end{figure}

The difficulty in solving \eqref{eq:problem_statement} lies in the size of the decision space $\C{X}$.
To overcome this issue and solve \eqref{eq:problem_statement}, we propose a data driven control strategy whose block scheme is represented in Figure \ref{fig:block_scheme} and described in the following.

\section{A data driven solution framework}
\label{sec:framework}

We assume the availability of a \emph{dataset} $\C{D} \subset \C{X}$, where each data sample, say $v_\R{d} \in \C{D}$, is a human velocity signal with known encoding level $\varepsilon(v_\R{d})$ of the target emotion.
Given a desired encoding level $e_\R{des}$, we further suppose that there exists a non-trivial partition $(\C{D}_{e_\R{des}}, \C{D}_{\neg e_\R{des}})$ of $\C{D}$, where
\begin{align}\label{eq:partition_dataset}
    \C{D}_{e_\R{des}} &\coloneqq \{v \in \C{D} \mid \varepsilon(v) = e_\R{des}\},\\
    \C{D}_{\neg e_\R{des}} &\coloneqq \{v \in \C{D} \mid \varepsilon(v) \ne e_\R{des}\}.
\end{align}

We then proceed according to the following steps.
\begin{enumerate}
\item We train a feedforward neural network to approximate the encoding function $\varepsilon$, using the labelled examples in $\C{D}$, and denote by $\hat{\varepsilon} : \C{X} \rightarrow \{0, 1\}$ the resulting approximation.

\item We choose the \emph{reference velocity signal}, say $v_r$, as the signal in the dataset such that
\begin{equation}\label{eq:reference_velocity}
    v_\R{r} = \R{proj}_{\C{D}_{e_\R{des}}}(v_\R{h}),
\end{equation}

\item We compute the output transformed velocity signal $v_a$ as the following combination of the velocity signal of the human participant $v_\R{h}$ and the reference velocity selected from the dataset. Specifically, we set:
\begin{equation}\label{eq:altered_motion}
    v_\R{a} = c v_\R{h} + (1 - c) v_\R{r},
\end{equation}
where  $c \in \C{C}$ is a \emph{blending coefficient} with $\C{C}$ being an appropriate discretization of the interval $[0, 1]$.

\item The position of the modified kinematics is then computed by inversion of \eqref{eq:relation_position_velocity}, enforcing $p_\R{a}(1) = p_\R{h}(1)$.
\end{enumerate}

Consider now the revised optimization problem
\begin{subequations}\label{eq:problem_statement_implemented}
    \begin{align}
        \max_{c \in \C{C}} \ \ & c,\\ 
        \text{s.t.} \ \
        & \hat{\varepsilon}(v_\R{a}) = e_\R{des}.\label{eq:constraint_emotion_implemented}
    \end{align}
\end{subequations}
In the next Lemma, we show that it is possible to solve 
\eqref{eq:problem_statement_implemented} in order to solve Problem \eqref{eq:problem_statement}, when \eqref{eq:altered_motion} is assumed.
    
\begin{lemma} \label{lem:equivalent_statements}
    Assume that the encoding function $\varepsilon$ is approximated by some $\hat{\varepsilon}$ such that
    \begin{equation}\label{eq:assumption_approximate_encoding}
    \forall v \in \C{X}, \quad
    \begin{dcases}
            \varepsilon(v) &\geq \hat{\varepsilon}(v) \quad \text{if } e_\R{des} = 1\\
            \varepsilon(v) &\leq \hat{\varepsilon}(v) \quad \text{if } e_\R{des} = 0.
    \end{dcases}
    \end{equation}
    Then, when $v_\R{a}$ is computed from \eqref{eq:altered_motion}, the solution $c^*$ to Problem \eqref{eq:problem_statement_implemented} yields a $v_\R{a}$ that is optimal for Problem \eqref{eq:problem_statement}.
\end{lemma}
\begin{proof}
Using \eqref{eq:altered_motion}, we have
\begin{equation}\label{eq:distance_implementation}
\begin{aligned}
    \R{dist}(v_\R{h}, v_\R{a}) &= \norm{v_\R{h} - v_\R{a}} = \norm{v_\R{h} - c v_\R{h} - (1-c)v_\R{r}}\\
    &= \norm{(1-c) (v_\R{h} - v_\R{r})} = \abs{1-c} \norm{(v_\R{h} - v_\R{r})}.
\end{aligned}
\end{equation}
Then, as  $v_\R{r}$ is selected according to \eqref{eq:reference_velocity}, from \eqref{eq:distance_implementation}, and recalling that $0 \le c \le 1$, it is immediate to see that minimizing $\R{dist}(v_\R{h}, v_\R{a})$  in \eqref{eq:problem_statement} corresponds to maximizing $c$.
Additionally, for a signal $v_\R{a}^*$ that satisfies \eqref{eq:constraint_emotion_implemented}, constraint \eqref{eq:constraint_encoding} holds as a consequence of \eqref{eq:assumption_approximate_encoding}.
Indeed, recall that $\varepsilon : \C{X} \rightarrow \{0, 1\}$; if $e_\R{des} = 1$, we have $\varepsilon(v_\R{a}^*) \geq \hat{\varepsilon}(v_\R{a}^*) = e_\R{des} = 1$ and thus $\varepsilon(v_\R{a}^*) = 1$; 
on the other hand, if $e_\R{des} = 0$, we obtain $\varepsilon(v_\R{a}^*) \le \hat{\varepsilon}(v_\R{a}^*) = e_\R{des} = 0$ and thus $\varepsilon(v_\R{a}^*) = 0$.
\end{proof}

\subsection{Offline solution}
\label{sec:heuristic_solution}

We start by proposing a heuristic solution to problem \eqref{eq:problem_statement_implemented} that can be used when the human movement kinematic signals to be altered have already been acquired and are available offline. 

Let $w : \{0, 1\} \times \C{X} \times \C{D}_{e_\R{des}} \rightarrow \C{C}$ be the (unknown) \emph{solution function} that, given some desired encoding level $e_\R{des}$, human velocity signal $v_\R{h}$, and reference velocity signal $v_\R{r}$, yields the solution $c^*$ to \eqref{eq:problem_statement_implemented}, i.e., such that $w : (e_\R{des}, v_\R{h}, v_\R{r}) \mapsto c^*$.

Define the \emph{restricted solution function} $\tilde{w} : \{0, 1\} \times \C{D}_{\neg e_\R{des}} \times \C{D}_{e_\R{des}} \rightarrow \C{C}$ as a restriction of $w$.
Note that both the domain and codomain of $\tilde{w}$ have finite cardinality; we assume  that they are small enough that $\tilde{w}$ can be computed by enumeration in a reasonable time.%
\footnote{If this is not the case, i.e., $\C{D}$ is relatively large, an approximator such as a feedforward neural network can be used to approximate $\tilde{w}$.}

Then, define the \emph{approximate solution function} $\hat{w} : \{0, 1\} \times \C{X} \times \C{D}_{e_\R{des}} \rightarrow \C{C}$ given by
\begin{equation}\label{eq:approx_static_sol}
    \hat{w}(e_\R{des}, v_\R{h}, v_\R{r}) \coloneqq \tilde{w}(e_\R{des}, \R{proj}_{\C{D}_{\neg e_\R{des}}}(v_\R{h}), v_\R{r}).
\end{equation}
which yields the suboptimal blending coefficient $\hat{c}^*$ to solve \eqref{eq:problem_statement} approximately.
Note that, unlike $w$, as will be show next $\hat{w}$ can be computed, because $\tilde{w}$ is known and $\R{proj}_{\C{D}_{\neg e_\R{des}}}$ can be computed.
A block scheme reviewing the strategy is reported in Figure \ref{fig:block_scheme}.

\subsection{Online computation}
\label{sec:online_computation}
Next, we propose a solution that generates the altered signal $v_\R{a}$ in real-time, as the human velocity signal $v_\R{h}$ is being measured.
To do so, we  introduce the {\em restriction} and {\em expansion} operators as follows.

For any \emph{reduced time duration} $\tau \in \{1, \dots, T\}$, let $\C{T}_\tau = \{1, \dots, \tau\}$ be a \emph{reduced time window}.
We let the \emph{$\tau$-restriction}, denoted by $\restr{\cdot}{\tau}$, be the restriction of a signal in $\C{X}$ to $\C{T}_\tau$.
By extension, for any $\C{S} \subseteq \C{X}$, $\restr{\C{S}}{\tau} \coloneqq \{ \restr{s}{\tau}\}_{s \in \C{S}}$ is the set of all signals in $\C{S}$ restricted to $\C{T}_\tau$.
Moreover, we define the \emph{$(\tau,\C{S})$-expansion} operator, denoted by $\R{expa}_{\tau}^{\C{S}} : \restr{\C{S}}{\tau} \rightarrow \C{S}$, yielding the inversion of $\tau$-restriction with respect to $\C{S}$.
Namely,
\begin{equation*}
    \forall \zeta \in \restr{\C{S}}{\tau}, \quad 
    \R{expa}_{\tau}^{\C{S}}(\zeta) = s \in \C{S} \text{ s.t. } \zeta = \restr{s}{\tau}.
\end{equation*}
To have $\R{expa}_{\tau}^{\C{S}}$ be well defined, it is required that, in $\C{S}$, there are no two elements whose $\tau$-restriction is the same, i.e., 
\begin{equation}\label{eq:well_posed_expa}
    \not\exists s_1, s_2 \in \C{S} : s_1 \ne s_2, \restr{s_1}{\tau} = \restr{s_2}{\tau}.
\end{equation}

\begin{figure}[t]
  \centering
  \includegraphics[width=0.45\linewidth]{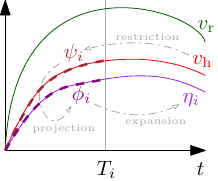}
  \caption{Example of the trajectories, projections, restrictions, and expansions described in Sec. \ref{sec:online_computation}.}
  \label{fig:projections_trajectories}
\end{figure}

We are now ready to present the online implementation of our proposed control strategy.
We assume the blending coefficient is time-varying, and with slight abuse of notation denote it as the function $c : \C{T} \rightarrow \C{C}$, and set $c(t) = 0$ for the first $T_0 < T$ time steps.
Define $\Delta T \in \BB{N}_{>1}$ and the time instants $T_i \coloneqq T_0 + i\Delta T$, $\forall i \in \C{I}$, where $\C{I} \coloneqq \{1, \dots, \lfloor \frac{T - T_0}{\Delta T} \rfloor \}$.
At each of these instants we measure the current available portion of the human velocity signal, that is $\psi_i \coloneqq \restr{v_\R{h}}{{T_i}}$ (see Figure \ref{fig:projections_trajectories}).
Next, we project $\psi_i$ onto the dataset, obtaining $\phi_i \coloneqq \R{proj}_{\restr{\C{D}_{\neg e_\R{des}}}{_{{T_i}}}}(\psi_i)$ (c.f.~\eqref{eq:partition_dataset}).
Finally, we recover the expansion of $\phi_i$ as
$\eta_i \coloneqq \R{expa}_{T_i}^{\C{D}_{\neg e_\R{des}}}(\phi_i)$.
Now, since $\eta_{i}\in \C{D}_{\neg e_\R{des}}$, it is possible to compute the blending coefficient as
\begin{equation} \label{eq:learned_blending_coefficient}
    c(t) = \tilde{w}(e_\R{des}, \eta_i, v_\R{r}), \quad \forall t \in \{T_{i-1} + 1, \dots, T_i\}. 
\end{equation}
In our case, to have \eqref{eq:well_posed_expa} and have $\R{expa}_{T_i}^{\C{D}_{\neg e_\R{des}}}$ be well defined, we make the following Assumption.
\begin{assumption}
    There do not exist different $s_1, s_2 \in \C{D}_{\neg e_\R{des}}$ such that $\restr{s_1}{{T_0}} = \restr{s_2}{{T_0}}$.
\end{assumption}

Moreover, since the whole $v_\R{h}$ is not available when the first segments of $v_\R{a}$ must be generated, it is not possible to use \eqref{eq:reference_velocity} to select the reference velocity signal $v_\R{r}$, and therefore we select it as $v_\R{r} = \R{proj}_{\C{D}_{e_\R{des}}}(\eta_0)$.

\section{Enforcing initial and terminal conditions} \label{sec:enforce_cond}
When a movement is carried out to perform some task, often some initial and/or terminal constraints on the modified kinematics must be fulfilled so as to match the initial and/or terminal conditions of the human movement. For instance,
when the movement is performed with the goal to reach an object, in Problem \eqref{eq:problem_statement} we have the additional constraint that the altered movement terminates sufficiently close to the human one.

Then, given the human velocity signal $v_\R{h}$, with position signal $p_{\R{h}}$, we extend Problem \eqref{eq:problem_statement_implemented} as
\begin{subequations}\label{eq:problem_statement_finalpos}
    \begin{align}
        \max_{c \in \C{C}} \ \ & c,\\ 
        \text{s.t.} \ \
        & \varepsilon(v_\R{a}) = e_\R{des},\label{eq:constr_emotion}\\
        & \norm{p_{\R{h}}(t_\R{term}(v_\R{h})) - p_{\R{a}}(t_\R{term}(v_\R{h}))} \leq \delta_{\R{pos}}, \label{eq:constr_finite_time_convergence}
    \end{align}
\end{subequations}%
where $\delta_{\R{pos}} \in \BB{R}_{>0}$ is a small threshold.
It is important to remark that, because of the new constraint \eqref{eq:constr_finite_time_convergence}, in principle there is no guarantee that there exists a solution to Problem \eqref{eq:problem_statement_finalpos}.
Hence, in practical scenarios it might be required to relax at least one of the constraints.

\subsection{Heuristic solution to the problem of reaching an object}

To provide a heuristic solution to Problem \eqref{eq:problem_statement_finalpos}, this time we generate the altered velocity signal as
\begin{equation} \label{eq:altered_motion_additive}   
    v_\R{a} = c v_\R{h} + (1 - c)v_\R{r} + v_u,
\end{equation}
with $v_u \in \C{X}$ being a correction term, given by
\begin{equation*}
    v_u = k_u \alpha \odot (p_\R{h} - p_\R{a}),
\end{equation*}
where $k_u \in \BB{R}_{>0}$ is a control gain,
$p_\R{h}(t) - p_\R{a}(t)$ can be understood as the integral error on velocity,
and $\alpha : \C{T} \rightarrow \{0, 1\}^\rho$ yields a vector of boolean variables, used to switch on or off the contribution of the correction term on specific degrees of freedom, and is computed via reinforcement learning.

In particular, following \cite{mnih2015human}, we use a Deep Q-Network approach (DQN), with state $\xi(t) = [p_\R{h}(t)\T\ p_\R{a}(t)\T\ v_\R{h}(t)\T\ v_\R{a}(t)\T]\T \in \BB{R}^{4\rho}$ and action $\alpha(t) \in \{0, 1\}^{\rho}$.
The reward function is defined according to the procedure described in \cite{de2023guaranteeing}, which allows to provide a guarantee for the satisfaction of \eqref{eq:constr_finite_time_convergence}, provided that the cumulative reward obtained is large enough.
Namely, let
\begin{equation*}
    g(t) \coloneqq \begin{dcases}
        1, & \text{if } \norm{p_\R{h}(t) - p_\R{a}(t)}_2 \leq \delta_\R{pos},\\
        0, & \text{otherwise}.
    \end{dcases}    
\end{equation*}
The reward at time $t$ is selected as 
\begin{equation}\label{eq:reward}
    r(t) = - k_r \norm{p_\R{a}(t) - p_\R{h}(t)}_2 - k_\alpha \norm{\alpha(t)}_\infty + r^\R{c}(t).
\end{equation}
where $k_r, k_\alpha \in \BB{R}_{>0}$ and
\begin{equation} \label{eq:reward_correction}
    r^{\R{c}}(t) = 
    \begin{cases}
        r_{\R{in}}^{\R{c}}, 
            &\text{if } g(t) = 1,\\
        r_{\R{exit}}^{\R{c}}, 
            &\text{if } g(t) = 0 \text{ and } g(t-1) = 1,\\
        0, & \text{otherwise}.
    \end{cases}
\end{equation}
where $r_{\R{in}}^{\R{c}}, r_{\R{exit}}^{\R{c}} \in \BB{R}$ are chosen following Algorithm 1 in \cite{de2023guaranteeing}.
In \eqref{eq:reward}, the term containing $\alpha$ is used to avoid applying the correction term $v_u$ when unnecessary.
The objective of the reinforcement learning model is to maximize $\sum_{t = 1}^T \gamma^t r(t)$ (see e.g., \cite{pmlr-v211-de-lellis23a}), where $\gamma \in [0, 1]$ is the \emph{discount factor}

During the learning phase, we run episodic tasks selecting each time a different $v_\R{h} \in \C{D}_{\neg e_\R{des}}$ and compute the reference profile $v_\R{r} \in \C{D}_{e_\R{des}}$ as well as the blending coefficient as described in Sec.
\ref{sec:heuristic_solution} \ref{sec:online_computation}.

\section{Validation} \label{sec:validation}

To validate our AI-based control architecture focused on encoding fear in human movements during a reach-to-grasp phase without actual fear, we define velocity signal $v$ with $\varepsilon(v) = 0$ indicating no fear and $\varepsilon(v) = 1$ indicating fear, setting $e_\R{des} = 1$. Our approach, detailed in Sec. \ref{sec:framework}, utilizes a dataset $\C{D}$ comprising velocity signals both with and without fear ($\C{D}_{e_\R{des}}$ and $\C{D}_{\neg e_\R{des}}$ respectively) for this purpose. 

For data collection, $n_\R{p} = 11$ naive participants were observed performing reach-to-grasp movements towards two sensorized cubes of identical size and weight but different colors (one blue, the other yellow) using an near infra-red optical motion capture system (Vicon Motion Systems Ltd, frame rate $100$ Hz). An unpleasant electrodermal stimulation was delivered upon touching one cube (e.g., the yellow one) in $33\%$ of the trials, with the identity of the cube causing stimulation counterbalanced among participants.%
\footnote{The experimental protocol has received approval by the ethical committee (Ethikkommission bei der Ärztekammer Hamburg).
All participants received written information about the purpose of the study and the electrodermal stimulation.
They were informed that they were allowed to withdraw from the experiment at any time.
Written informed consent was obtained by all participants prior to the experimental session.
Electrodermal stimulation was well tolerated by participants. 
No discomfort or adverse effects were reported by participants or noticed by the experimenter during the calibration procedure, nor during or after the experimental session.}
The recorded movement trajectories were classified as not encoding fear ($\C{D}_{\neg e_\R{des}}$) when participants reached for a cube knowing they would not receive any stimulation, and as encoding fear ($\C{D}_{e_\R{des}}$) when they performed reaching movements with the expectation that they may receive stimulation, based on knowledge gained from previous experiences.

The resulting dataset $\C{D}$ consists of 458 samples, with 197 labeled as encoding fear and 261 as not encoding fear. Each sample records the 3D velocity and position of a participant's wrist at a 100 Hz sampling rate. 

\subsection{Training of the approximate encoding function $\hat{\varepsilon}$}
\label{sec:training_approximate_encoding_function}
\begin{figure}[t]
  \begin{tikzpicture}
    \node[anchor=south west,inner sep=0] (image) at (0,0) {\includegraphics[width=1.05\linewidth]{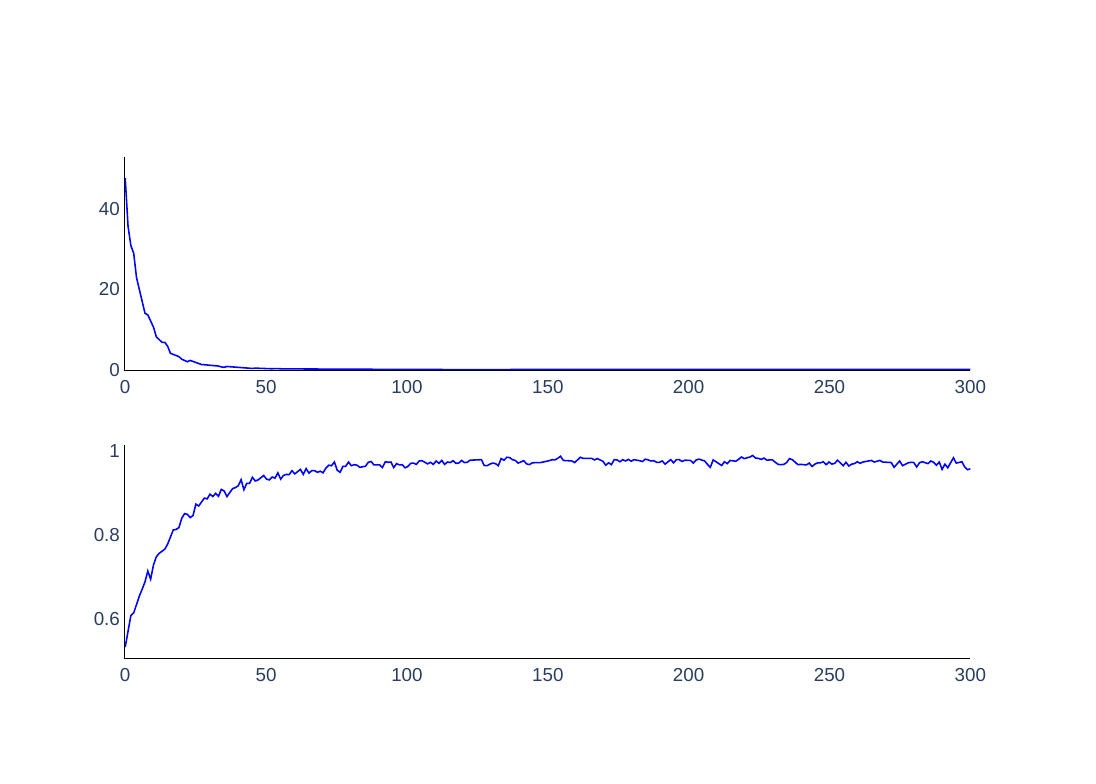}};
    \begin{scope}[x={(image.south east)},y={(image.north west)}]
    \node at (0.5,0.05) {epochs};
    \node[rotate=90]  at (0.05,0.65) {loss};
    \node[rotate=90]  at (0.05,0.3) {accuracy};
    \end{scope}
  \end{tikzpicture}
  \caption{Relevant quantities concerning the training of the approximate encoding function (see Sec. \ref{sec:training_approximate_encoding_function}).
  Results are computed as an average of the 5 sessions used in cross validation.}
  \label{fig:neural_training}
\end{figure}
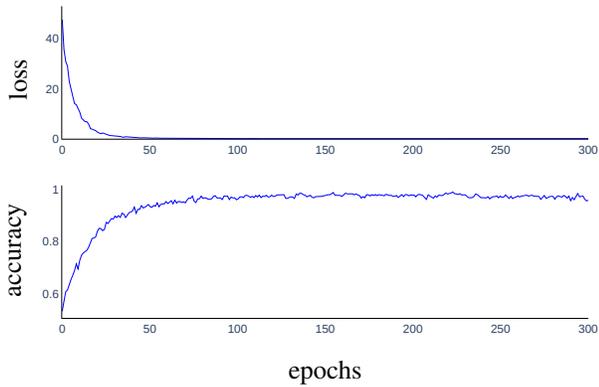

To compute the restricted solution function \(\tilde{w}\) as discussed in Sec. \ref{sec:heuristic_solution}, we train a neural classifier to approximate the encoding function \(\varepsilon\), where the classifier inputs a velocity signal \(v \in \C{X}\) and outputs its encoding level \(\varepsilon(v)\). We use a feedforward neural network with an input layer of 60 nodes (representing the 20 points per each axis of the velocity signal available), a hidden layer of 200 neurons with ReLU activation functions, and a sigmoid-activated output layer. Input signals are resampled to 20 points per axis after restriction and cubic interpolation. Training employs the Adam optimizer with a learning rate of \(0.001\) and dropout on the hidden nodes to prevent overfitting. To fully validate, we perform a 5-fold cross-validation with a 70-30 split for training and validation, respectively. Training average perfomances are depicted in Figure \ref{fig:neural_training}.

To minimize misclassification's impact, we apply a clipping layer on the classifier's output, categorizing signals as encoding fear for \(\varepsilon(v) \in (0.9, 1]\), not encoding fear for \(\varepsilon(v) \in [0, 0.1)\), and discarding unclassified signals with \(\varepsilon(v) \in [0.1, 0.9]\). 
The model achieves 87\% accuracy for fear signals and 83\% for non-fear signals on the validation dataset, with misclassifications at minimum with 3\% and 1.5\% respectively.

Post-validation, the network with the highest accuracy on the validation set is selected, identifying two specific sets for final evaluation: \(\C{D}_{\neg e_\R{des}}^{\R{val}}\) with 65 samples not encoding fear and \(\C{D}_{e_\R{des}}^{\R{val}}\) with 73 samples encoding fear.

\subsection{Computation of $\tilde{w}$ and enforcement of the terminal conditions}
\label{sec:computation_omega_and_RL}

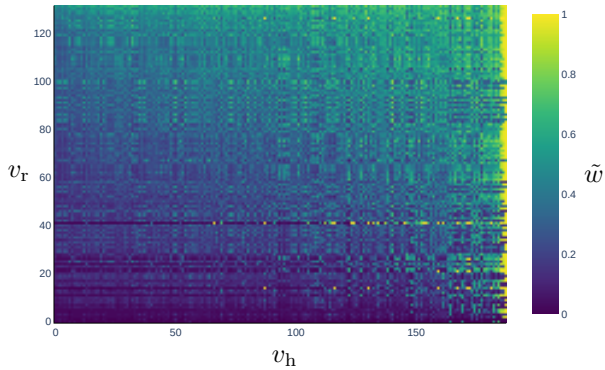
\begin{figure}[t]
  \begin{tikzpicture}
    \node[anchor=south west,inner sep=0] (image) at (0,0) {\includegraphics[width=0.95\linewidth]{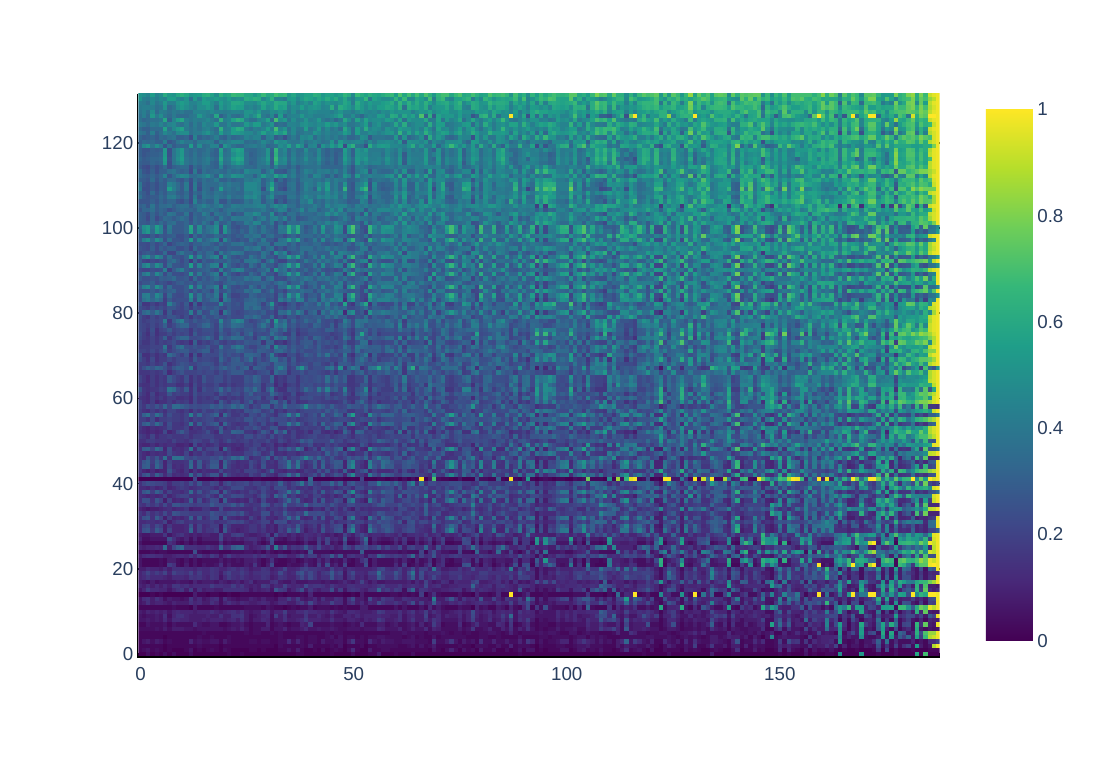}};
    \begin{scope}[x={(image.south east)},y={(image.north west)}]
      \node at (0.5,0.08) {$v_\R{h}$};
      \node at (0.07,0.5) {$v_\R{r}$};
      \node at (1,0.5) {$\tilde{w}$};
    \end{scope}
  \end{tikzpicture}
  \caption{Values of the restricted solution function $\tilde{w}(1, v_\R{h}, v_\R{r})$ for $v_\R{h} \in \C{D}_{\neg e_\R{des}}$ and $v_\R{r} \in \C{D}_{e_\R{des}}$ (see Sec. \ref{sec:heuristic_solution}).
  }
  \label{fig:blending_table}
\end{figure}

Next, we compute $\tilde{w}$ exhaustively (see Sec. \ref{sec:heuristic_solution}).
To do so, we set $\C{C}$ to be the discretization of $[0, 1]$, with 50 equally spaced values, including $0$ and $1$, letting $\Delta c \coloneqq 1/50$, and follow Algorithm 
\ref{alg:computation_restricted_solution_function}.
Figure \ref{fig:blending_table} illustrates the computed values of $\tilde{w}$ for all velocity signals in the dataset.

\begin{algorithm}[t]
  \caption{Computation of restricted solution function}
  \label{alg:computation_restricted_solution_function}
  \KwIn{Datasets $\C{D}_{e_\R{des}}$ and $\C{D}_{\neg e_\R{des}}$; $\C{C} = \{0, \Delta c, 2 \Delta c, \dots, 1\}$; approximate encoding function $\hat{\varepsilon}$.}%
  \KwOut{Restricted solution function $\tilde{w}$.}
  %
  \BlankLine
  \For{$v_\R{ne} \in \C{D}_{\neg e_\R{des}}$}{
      \For{$v_\R{r} \in \C{D}_{e_\R{des}}$}{
      $c \leftarrow 1$; \quad \texttt{is\textunderscore done} $\leftarrow$ false\;
      \While{$\neg$\texttt{is\textunderscore done}}{
        $v_\R{a} \leftarrow c v_\R{ne} + (1-c)v_\R{r}$%
        \tcp*{Exploiting \eqref{eq:altered_motion}}
        \If{$\hat{\varepsilon}(v_\R{a}) = e_\R{des} \ \vee \ c = 0$\tcp*{$e_\R{des} = 1$}}{
            \texttt{is\textunderscore done} $\leftarrow$ true; \quad 
            $\tilde{w}(1, v_\R{ne}, v_\R{r}) \leftarrow c$\;
        }
        \lElse{$c \leftarrow c - \Delta c$}
      }
    }
  }
\end{algorithm}

\begin{figure}[t]
  \begin{tikzpicture}
    \node[anchor=south west,inner sep=0] (image) at (0,0) {\includegraphics[width=1.05\linewidth]{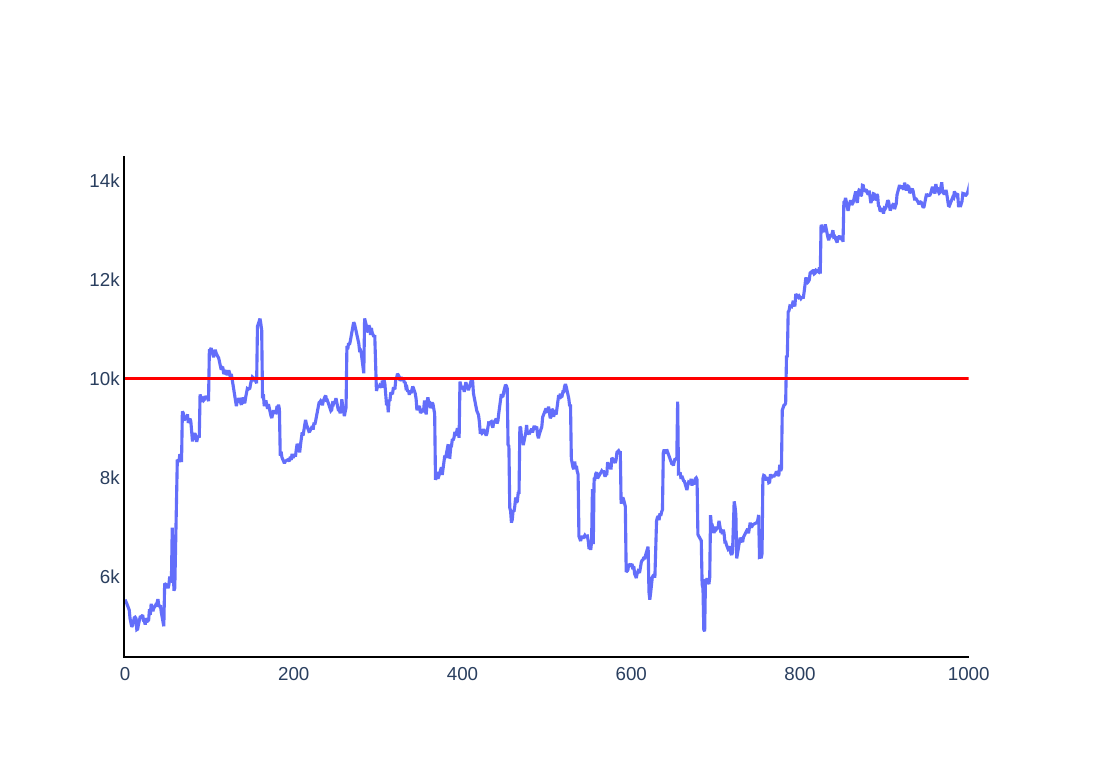}};
    \begin{scope}[x={(image.south east)},y={(image.north west)}]
    \node at (0.5,0.05) {episodes};
    \node[rotate=90]  at (0.05,0.5) {cumulative discounted reward};
    \end{scope}
  \end{tikzpicture}
  \caption{Moving average of 100-sample cumulative discounted rewards per episode.
  Red line: threshold value $\sigma =  10000$, defined in \cite{de2023guaranteeing}. The agent surpasses the threshold after 800 episodes, indicating successful constraint enforcement \eqref{eq:constr_finite_time_convergence} on training examples in $\C{D}_{\neg e_\R{des}}$.}
  \label{fig:rl_training}
\end{figure}

To train the artificial agent described in Sec. \ref{sec:enforce_cond}, we run $E = 1100$ episodes, each lasting $200$ steps with a DQN algorithm. The DQN approach used consists in  a Deep Neural Network made of two hidden layer of $128$ nodes with ReLU activation function. Moreover, at the end of each episode we copy the weights of the neural approximator in a second target network with same structure. We select the discount factor $\gamma = 0.99$, learning rate $0.001$ and set random exploration probability to $1$ with a decreasing factor of $0.995$ applied at the end of each episode. Moreover, the reward \eqref{eq:reward} uses coefficients $k_\R{r} = -0.01$, $k_\R{\alpha} = -10$ and the correction term \eqref{eq:reward_correction} is shaped to enforce constraint \eqref{eq:constr_finite_time_convergence} with $\delta_\R{pos} = 20\,\text{mm}$. Consequently, according to \eqref{eq:constr_finite_time_convergence}, the  evaluation of the fulfillment of condition \eqref{eq:constr_finite_time_convergence} is encoded in the discounted cumulative rewards, as explained in \cite{de2023guaranteeing}.
The results of this training procedure is schematically depicted in Figure \ref{fig:rl_training}.

\subsection{Validation of the control strategy}
We validate the online control strategy introduced in Sec. \ref{sec:online_computation} using the validation set \(\C{D}_{\neg e_\R{des}}^{\R{val}}\). The numerical simulation run with a sampling time of $0.01\,\text{s}$ in accordance with the sampling rate of the motion capture system (see Sec. \ref{sec:validation}).

\begin{figure}[t]
  \begin{tikzpicture}
    \node[anchor=south west,inner sep=0] (image) at (0,0) {\includegraphics[width=1.05\linewidth]{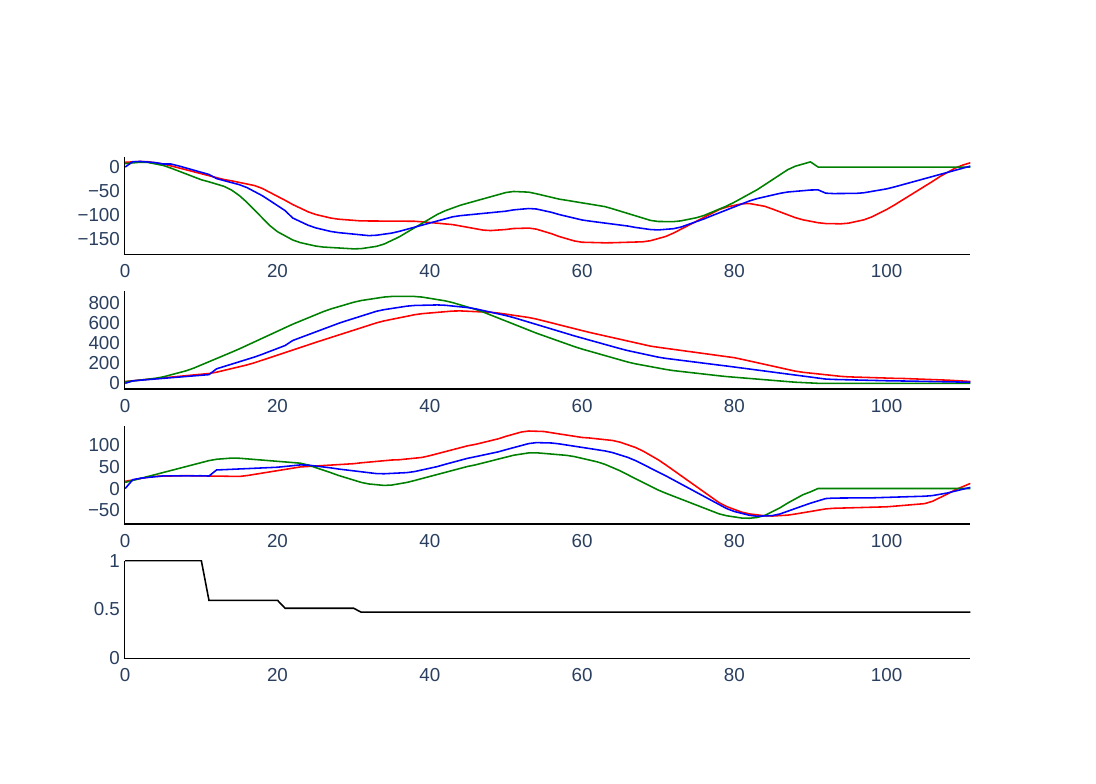}};
    \begin{scope}[x={(image.south east)},y={(image.north west)}]
      \node at (0.5,0.05) {samples (sampling time $10\,\text{ms}$)};
      \node at (0.05,0.72) {$x$};
      \node at (0.05,0.56) {$y$};
      \node at (0.05,0.4) {$z$};
      \node at (0.05,0.22) {$c(t)$};
    \end{scope}
  \end{tikzpicture}
  \caption{Representative online alteration of a human motion signal in the 3D space, whose components are denoted as $x$, $y$, $z$.
  The red line is $v_\R{h}$, the green line is $v_\R{r}$, and the blue line is $v_\R{a}$.}
  \label{fig:solution1}
\end{figure}
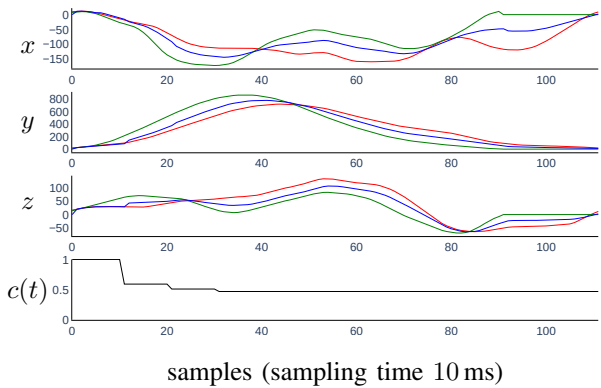

In Figure \ref{fig:solution1}, we showcase an example where a human velocity signal \(v_\R{h}\), not encoding fear, is transformed into \(v_\R{a}\) that encodes fear, as per the approximate encoding function \(\hat{\varepsilon}\). A more extensive validation, omitted here for the sake of brevity, demonstrated that our approach, when applied to each of the 65 experimentally obtained human movements in \(\C{D}_{\neg e_\R{des}}^{\R{val}}\) using a blending coefficient computed online, is able to achieve an $89\%$ success rate, with the constraint on the final position (not enforced in this case) being satisfied in $24\%$ of samples.

Applying solution \eqref{eq:altered_motion_additive} with an artificially added action, as outlined in Sec. \ref{sec:computation_omega_and_RL}, increased the final position constraint satisfaction to $90\%$. However, the success rate for classifying ``fear'' decreased to $65\%$. This reduction is expected due to the alteration process being modified by the trained agent's additive action, necessitating a trade-off between the two objectives. The decline in fear classification success is also attributed to the conservative assumptions of the neural classifier discussed in Sec.~\ref{sec:training_approximate_encoding_function}.

The validation results presented above confirm the effectiveness of the proposed approach as a promising solution to encode information such as desired emotional states in the kinematics of avatars or robots. This can be instrumental to enhance their social interaction with humans when performing joint tasks, an aspect we are currently investigating in our ongoing work.

\addtolength{\textheight}{-12cm}   







\begin{thebibliography}{10}
\providecommand{\url}[1]{#1}
\csname url@samestyle\endcsname
\providecommand{\newblock}{\relax}
\providecommand{\bibinfo}[2]{#2}
\providecommand{\BIBentrySTDinterwordspacing}{\spaceskip=0pt\relax}
\providecommand{\BIBentryALTinterwordstretchfactor}{4}
\providecommand{\BIBentryALTinterwordspacing}{\spaceskip=\fontdimen2\font plus
\BIBentryALTinterwordstretchfactor\fontdimen3\font minus \fontdimen4\font\relax}
\providecommand{\BIBforeignlanguage}[2]{{%
\expandafter\ifx\csname l@#1\endcsname\relax
\typeout{** WARNING: IEEEtran.bst: No hyphenation pattern has been}%
\typeout{** loaded for the language `#1'. Using the pattern for}%
\typeout{** the default language instead.}%
\else
\language=\csname l@#1\endcsname
\fi
#2}}
\providecommand{\BIBdecl}{\relax}
\BIBdecl

\bibitem{becchio2018seeing}
C.~Becchio, A.~Koul, C.~Ansuini, C.~Bertone, and A.~Cavallo, ``Seeing mental states: An experimental strategy for measuring the observability of other minds,'' \emph{Physics of life reviews}, vol.~24, pp. 67--80, 2018.

\bibitem{podda2017heaviness}
J.~Podda, C.~Ansuini, R.~Vastano, A.~Cavallo, and C.~Becchio, ``The heaviness of invisible objects: Predictive weight judgments from observed real and pantomimed grasps,'' \emph{Cognition}, vol. 168, pp. 140--145, 2017.

\bibitem{scaliti_kinematic_2023}
E.~Scaliti, K.~Pullar, G.~Borghini, A.~Cavallo, S.~Panzeri, and C.~Becchio, ``Kinematic priming of action predictions,'' \emph{Current Biology}, vol.~33, no.~13, pp. 2717--2727.e6, 2023.

\bibitem{gallivan2018decision}
J.~P. Gallivan, C.~S. Chapman, D.~M. Wolpert, and J.~R. Flanagan, ``Decision-making in sensorimotor control,'' \emph{Nature Reviews Neuroscience}, vol.~19, no.~9, pp. 519--534, 2018.

\bibitem{wispinski2020models}
N.~J. Wispinski, J.~P. Gallivan, and C.~S. Chapman, ``Models, movements, and minds: bridging the gap between decision making and action,'' \emph{Annals of the New York Academy of Sciences}, vol. 1464, no.~1, pp. 30--51, 2020.

\bibitem{montobbio2022intersecting}
N.~Montobbio, A.~Cavallo, D.~Albergo, C.~Ansuini, F.~Battaglia, J.~Podda, L.~Nobili, S.~Panzeri, and C.~Becchio, ``Intersecting kinematic encoding and readout of intention in autism,'' in \emph{Proceedings of the National Academy of Sciences}, vol. 119, no.~5.\hskip 1em plus 0.5em minus 0.4em\relax National Acad Sciences, 2022, p. e2114648119.

\bibitem{mcellin2018identifying}
L.~McEllin, N.~Sebanz, and G.~Knoblich, ``Identifying others’ informative intentions from movement kinematics,'' \emph{Cognition}, vol. 180, pp. 246--258, 2018.

\bibitem{strachan2021evaluating}
J.~W. Strachan, A.~Curioni, M.~D. Constable, G.~Knoblich, and M.~Charbonneau, ``Evaluating the relative contributions of copying and reconstruction processes in cultural transmission episodes,'' \emph{Plos one}, vol.~16, no.~9, p. e0256901, 2021.

\bibitem{mortl2014rhythm}
A.~M{\"o}rtl, T.~Lorenz, and S.~Hirche, ``Rhythm patterns interaction-synchronization behavior for human-robot joint action,'' \emph{PloS one}, vol.~9, no.~4, p. e95195, 2014.

\bibitem{huang2021connecting}
X.~Huang, W.~Wu, and H.~Qiao, ``Connecting model-based and model-free control with emotion modulation in learning systems,'' \emph{IEEE Transactions on Systems, Man, and Cybernetics: Systems}, vol.~51, no.~8, pp. 4624--4638, 2021.

\bibitem{huang2018braininspired}
X.~Huang, W.~Wu, H.~Qiao, and Y.~Ji, ``Brain-inspired motion learning in recurrent neural network with emotion modulation,'' \emph{IEEE Transactions on Cognitive and Developmental Systems}, vol.~10, no.~4, pp. 1153--1164, 2018.

\bibitem{claret2017exploiting}
J.-A. Claret, G.~Venture, and L.~Basa{\~n}ez, ``Exploiting the robot kinematic redundancy for emotion conveyance to humans as a lower priority task,'' \emph{International Journal of Social Robotics}, vol.~9, no.~2, pp. 277--292, 2017.

\bibitem{lourens2010communicating}
T.~Lourens, R.~Van~Berkel, and E.~Barakova, ``Communicating emotions and mental states to robots in a real time parallel framework using {{Laban}} movement analysis,'' \emph{Robotics and Autonomous Systems}, vol.~58, no.~12, pp. 1256--1265, 2010.

\bibitem{bernardet2019assessing}
U.~Bernardet, S.~Fdili~Alaoui, K.~Studd, K.~Bradley, P.~Pasquier, and T.~Schiphorst, ``Assessing the reliability of the laban movement analysis system,'' \emph{PloS one}, vol.~14, no.~6, p. e0218179, 2019.

\bibitem{wu2022robotic}
K.~Wu, L.~Chen, K.~Wang, M.~Wu, W.~Pedrycz, and K.~Hirota, ``Robotic arm trajectory generation based on emotion and kinematic feature,'' in \emph{2022 {{International Power Electronics Conference}} ({{IPEC-Himeji}} 2022- {{ECCE Asia}})}, 2022, pp. 1332--1336.

\bibitem{lombardi2021dynamic}
M.~Lombardi, D.~Liuzza, and M.~di~Bernardo, ``Dynamic input deep learning control of artificial avatars in a multi-agent joint motor task,'' \emph{Frontiers in Robotics and AI}, vol.~8, 2021.

\bibitem{atkinson_emotion_2004}
A.~Atkinson, W.~Dittrich, A.~Gemmell, and A.~Young, ``Emotion perception from dynamic and static body expressions in point-light and full-light displays,'' \emph{Perception}, vol.~33, no.~6, pp. 717--46, 2004.

\bibitem{llobera2022playing}
J.~Llobera, V.~Jacquat, C.~Calabrese, and C.~Charbonnier, ``Playing the mirror game in virtual reality with an autonomous character,'' \emph{Scientific Reports}, vol.~12, no.~1, p. 21329, 2022.

\bibitem{melzer_how_2019}
A.~Melzer, T.~Shafir, and R.~P. Tsachor, ``How do we recognize emotion from movement? specific motor components contribute to the recognition of each emotion,'' \emph{Frontiers in Psychology}, vol.~10, 2019.

\bibitem{10.3389/frobt.2020.532279}
M.~Spezialetti, G.~Placidi, and S.~Rossi, ``Emotion recognition for human-robot interaction: Recent advances and future perspectives,'' \emph{Frontiers in Robotics and AI}, vol.~7, 2020.

\bibitem{turri_decoding_2022}
G.~Turri, A.~Cavallo, L.~Romeo, M.~Pontil, A.~Sanfey, S.~Panzeri, and C.~Becchio, ``Decoding social decisions from movement kinematics,'' \emph{iScience}, vol.~25, no.~12, p. 105550, 2022.

\bibitem{mnih2015human}
V.~Mnih, K.~Kavukcuoglu, D.~Silver, A.~A. Rusu, J.~Veness, M.~G. Bellemare, A.~Graves, M.~Riedmiller, A.~K. Fidjeland, G.~Ostrovski \emph{et~al.}, ``Human-level control through deep reinforcement learning,'' \emph{Nature}, vol. 518, no. 7540, pp. 529--533, 2015.

\bibitem{de2023guaranteeing}
F.~De~Lellis, M.~Coraggio, G.~Russo, M.~Musolesi, and M.~di~Bernardo, ``Guaranteeing control requirements via reward shaping in reinforcement learning,'' \emph{arXiv preprint arXiv:2311.10026}, 2023.

\bibitem{pmlr-v211-de-lellis23a}
F.~{De Lellis}, M.~Coraggio, G.~Russo, M.~Musolesi, and M.~di~Bernardo, ``{CT-DQN}: Control-tutored deep reinforcement learning,'' in \emph{Proceedings of The 5th Annual Learning for Dynamics and Control Conference (L4DC 2023)}, vol. 211.\hskip 1em plus 0.5em minus 0.4em\relax PMLR, 2023, pp. 941--953.

\end{thebibliography}
\end{document}